\documentclass[fleqn,usenatbib]{mnras}

\usepackage{newtxtext,newtxmath}

\usepackage[T1]{fontenc}

\DeclareRobustCommand{\VAN}[3]{#2}
\let\VANthebibliography\thebibliography
\def\thebibliography{\DeclareRobustCommand{\VAN}[3]{##3}\VANthebibliography}

\usepackage{graphicx}	
\usepackage{amsmath}	
\usepackage[utf8]{inputenc}
\usepackage{ulem}

\title[Splashback radius around X-ray galaxy clusters]{The eROSITA Final Equatorial-Depth Survey (eFEDS) - Splashback radius of X-ray galaxy clusters using galaxies from HSC survey}

\author[Rana et al.]{
Divya Rana$^{1}$\thanks{E-mail: divyar@iucaa.in},
Surhud More$^{1,2}$\thanks{E-mail: surhud@iucaa.in},
Hironao Miyatake$^{2,3,4}$,
Sebastian Grandis$^{5,6}$,
Matthias Klein$^{6}$,
\newauthor{
Esra Bulbul$^{7}$,
I-Non Chiu$^{8,9,10}$,
Satoshi Miyazaki$^{11}$,
Neta Bahcall$^{12}$
}
\\
$^{1}$ Inter University Centre for Astronomy and Astrophysics, Ganeshkhind, Pune 411007, India\\
$^{2}$ Kavli Institute for the Physics and Mathematics of the Universe (WPI), University of Tokyo, 5-1-5, Kashiwanoha, 2778583, Japan\\
$^{3}$ Kobayashi-Maskawa Institute for the Origin of Particles and the Universe (KMI), Nagoya University, Nagoya, 464-8602, Japan\\
$^{4}$ Institute for Advanced Research, Nagoya University, Nagoya 464-8601, Japan\\
$^{5}$ Institut f\"ur Astro- und Teilchenphysik, Universit\"at Innsbruck, Technikerstr. 25/8, 6020 Innsbruck, Austria\\
$^{6}$ Faculty of Physics, Ludwig-Maximilians-Universität, Scheinerstr. 1,
81679, Munich, Germany\\
$^{7}$ Max-Planck-Institut für extraterrestrische Physik, Gießenbachstraße 1, D-85748 Garching, Germany\\
$^{8}$ Department of Physics, National Cheng Kung University, 70101 Tainan, Taiwan\\
$^{9}$ Department of Astronomy, School of Physics and Astronomy, and Shanghai Key Laboratory for Particle Physics and Cosmology,\\
\,\,\,\,Shanghai Jiao Tong University, Shanghai 200240, China\\
$^{10}$ Tsung-Dao Lee Institute, Shanghai Jiao Tong University, Shanghai 200240, China\\
$^{11}$ Subaru Telescope, National Astronomical Observatory of Japan, 650, N Aohoku Place, Hilo, HI 96720 USA\\
$^{12}$ Department of Astrophysical Sciences, Peyton Hall, 4 Ivy Lane, Princeton University, Princeton, NJ 08544
}

\date{Accepted XXX. Received YYY; in original form ZZZ}

\pubyear{2023}

\begin{document}
\label{firstpage}
\pagerange{\pageref{firstpage}--\pageref{lastpage}}
\maketitle

\begin{abstract}
We present the splashback radius measurements around the SRG/eROSITA eFEDS X-ray selected galaxy clusters by cross-correlating them with HSC S19A photometric galaxies. The X-ray selection is expected to be less affected by systematics related to projection that affects optical cluster finder algorithms. We use a nearly volume-limited sample of 109 galaxy clusters selected in 0.5-2.0 keV band having luminosity $L_X > 10^{43.5}\,{\rm erg s^{-1} h^{-2}}$ within the redshift $z<0.75$ and obtain measurements of the projected cross-correlation with a signal-to-noise of $17.43$. We model our measurements to infer a three-dimensional profile and find that the steepest slope is sharper than $-3$ and associate the location with the splashback radius. We infer the value of the 3D splashback radius $r_{\rm sp} = 1.45^{+0.30}_{-0.26}\,{\rm h^{-1} Mpc}$. We also measure the weak lensing signal of the galaxy clusters and obtain halo mass $\log[M_{\rm 200m}/{\rm h^{-1}M_\odot}] = 14.52 \pm 0.06$ using the HSC-S16A shape catalogue data at the median redshift $z=0.46$ of our cluster sample. We compare our $r_{\rm sp}$ values with the spherical overdensity boundary $r_{\rm 200m} = 1.75 \pm 0.08\,{\rm h^{-1} Mpc}$ based on the halo mass which is consistent within $1.2\sigma$ with the $\Lambda$CDM predictions. Our constraints on the splashback radius, although broad, are the best measurements thus far obtained for an X-ray selected galaxy cluster sample.
\end{abstract}

\begin{keywords}
galaxies: clusters: general -- cosmology: observations -- (cosmology:) large-scale structure of Universe
\end{keywords}



\section{Introduction}
The gravitational collapse of high-density peaks in the initial matter distribution results in the formation of virialized massive dark matter halos in the Universe. The most massive of these halos host clusters of galaxies that we see today \citep[see][for a recent review]{2012ARA&A..50..353K,2019SSRv..215....7W, 2020NatRP...2...42V}. The matter distribution in the dark matter halos is driven by the profile of the initial density peak from which it forms. There have been extensive theoretical and numerical studies to understand the structure of the dark matter halos \citep[e.g.][]{1972ApJ...176....1G, 1984ApJ...281....1F, 1985ApJS...58...39B, 1997ApJ...490..493N, 1999ApJ...524L..19M}. Since the seminal study of \citep{1997ApJ...490..493N}, it has been well known that dark matter halos follow self-similar density distribution in their inner regions, known as the Navarro-Frenk-White (NFW) profile. 

On the other hand, the outskirts of massive dark matter halos have received attention in the last few years and both theoretical and observational aspects of it remain a subject of active research. \citet[][]{2014ApJ...789....1D} showed that the matter distribution at the outskirts of stacked dark matter halo profiles differs from extrapolations of the NFW profile. In the outskirts of these halos, the density distribution shows logarithmic slopes ($d\log\rho/d\log r$) which are much steeper than the asymptotic value of $-3$ expected from the NFW profile. The location of the steepest slope also is dependent upon the mass accretion rate of these dark matter halos. \citet{2014JCAP...11..019A} explicitly showed with a phase space analysis that these locations correspond to the position where recently accreted particles reach the apocenters of the orbits for the first time. 

The resultant sharp drop in the density profile at this location is reminiscent of the last density caustic predicted in the model of secondary infall around a spherically symmetric overdensity \citep[][]{1984ApJ...281....1F,1985ApJS...58...39B,2016MNRAS.459.3711S}. \citet{2014JCAP...11..019A} termed this location the "splashback radius" as after reaching the apocenters, the particles are expected to splash back in to the halo. Subsequently, the splashback radius has been proposed as a physical boundary of the dark matter halo and found to be primarily dependent on the mass of the halo, its accretion rate and redshift \citep[][]{2015ApJ...810...36M, 2017ApJ...843..140D}. These results triggered an interest in various aspects of splashback radius studies in simulations \citep[see e.g ][]{2017ApJ...841...34M,2018PhRvD..98b3523O, 2018MNRAS.478.5366F, 2020MNRAS.493.4763M, 2020MNRAS.493.2765S, 2020MNRAS.499.3534X,2020ApJ...903...87D, 2021MNRAS.500.4181D, 2022MNRAS.513..835O}, for different dark matter and dark energy theories \citep{2018JCAP...11..033A, 2019PhRvD..99f4030C, 2020JCAP...02..024B} and galaxy evolution scenarios in clusters \citep[][]{2022MNRAS.512.4378D}.

Observational investigations of the splashback radius have focussed on galaxy clusters as the density drops at the splashback radius are expected to be quite significant due to their higher current accretion rates. Secondly on cluster scales it is easier to select isolated halos than on galaxy and group scales. \citet[][]{2016ApJ...825...39M} used the cross-correlation of the SDSS redMaPPer clusters \citep[][]{2014ApJ...785..104R} with SDSS photometric galaxies to present the first detection of the splashback radius. They detected a steepening of the projected galaxy number density profile as observed by \citet{2014ApJ...789....1D} in numerical simulations. However, the inferred location of the splashback radius was found to be about 20 percent smaller than expected from the dark matter simulations based on the $\Lambda{\rm CDM}$ model \citep[][]{2015ApJ...810...36M}. The robustness of these results to effects such as the miscentering of clusters and to priors were demonstrated in \citet{2017ApJ...841...18B}. \citet{2017MNRAS.470.4767B} argued that selection effects induced by optical clusters could affect the cross-correlation measurements and could therefore be important in understanding the origin of this difference. Using clusters selected by a mock redMaPPer algorithm run on a simulation, \citet{2019MNRAS.490.4945S} showed that these differences could indeed arise from such optical selection effects. Such selection effects are quite sensitive to the background subtraction scheme employed while running the cluster finding algorithm. \citet[][]{2020PASJ...72...64M} explored clusters selected with the CAMIRA cluster finder \citep[][]{2014MNRAS.444..147O} using data from the Subaru Hyper Suprime-Cam survey \citep{2018PASJ...70S...4A} which use a local background subtraction scheme. They obtained measurements of the splashback radius which were consistent with $\Lambda{\rm CDM}$ prediction.       

Such issues related to the projection effects in optically selected galaxy clusters can be avoided by using clusters selected using the Sunyaev Zeldovich (SZ) effect \citep{1970Ap&SS...7....3S, 1980MNRAS.190..413S}. \citet{2019ApJ...874..184Z}  used galaxy clusters selected from the Planck SZ survey and cross-correlated them with galaxies detected in the Pan-STARRS. \citet{2019MNRAS.487.2900S} used SZ clusters selected from the Atacama
Cosmology Telescope (ACT) Polarimeter \citep[][]{2018ApJS..235...20H} and the South Pole Telescope  \citep[SPT,][]{2015ApJS..216...27B} SZ survey along with galaxy catalog data from the Dark Energy Survey \citep[DES,][]{2005astro.ph.10346T}. They use both the galaxy number density and weak lensing profiles to obtain the constraints on the splashback radius and found the results consistent with the $\Lambda{\rm CDM}$ predictions, albeit with a larger errorbar which does not entirely preclude the initial results from optically selected clusters. Similarly, \citet[][]{2021MNRAS.507.5758S} show consistency of the measured splashback radius with expectations from $\Lambda{\rm CDM}$ using a larger cluster catalogue with improved precision, yet not with errors that could rival those obtained with optical clusters \citep[][]{2016ApJ...825...39M,2017ApJ...841...18B, 2018ApJ...864...83C, 2020PASJ...72...64M}. Furthermore, studies along the same lines have used galaxy clusters identified using the X-rays emitted by the bremsstrahlung emission from ICM \citep{2017ApJ...836..231U, 2019MNRAS.485..408C, 2021ApJ...911..136B} and found similar results.

In this study, we use galaxy clusters selected based on their X-ray emission and cross-correlate their positions with those of optical galaxies \citep{2016ApJ...825...39M, 2017ApJ...841...18B, 2018ApJ...864...83C, 2020PASJ...72...64M}. Previous studies with X-ray clusters \citep{2017ApJ...836..231U, 2019MNRAS.485..408C, 2021ApJ...911..136B} are based on either a limited number of galaxy clusters due to the shallower depth of the survey or their limited area coverage. Instead, our goal is to use the X-ray galaxy clusters from the extended ROentgen Survey with an Imaging Telescope Array  \citep[eROSITA,][]{2012arXiv1209.3114M, 2021A&A...647A...1P} on board the SRG mission, a highly sensitive space-based X-ray telescope launched in July 2019. The eROSITA mission aims to conduct an all sky survey in X-rays which will yield a catalog of $\sim 100000$ X-ray galaxy clusters by the end of four years of operation \citep{2014_Borm}. Before observing the entire sky, eROSITA first collected data from a smaller equatorial field of approximately 140 sq.~deg.~at its planned depth to test its performance. The X-ray galaxy cluster data provided by the eROSITA final equatorial depth survey (eFEDS) has been extensively studied for cluster cosmology \citep[see for e.g. ][]{2022_sanders, 2022_ramos, 2022_ghirardini, 2022_chiu,2022_Chiu_b, 2022_bulbul, 2022_Bahar,  2022A&A...661A...4K}. As a pilot study we will use the eFEDS X-ray galaxy clusters and obtain constraints on the location of the splashback radius by measuring the number density profiles of galaxies correlated with these clusters. 

Inference of the splashback radius using measurements of the galaxy number density profile is possible if galaxies act as test particles in the cluster potential and dynamically are distributed similar to dark matter particles. Galaxies residing in massive subhaloes are expected to be affected by dynamical friction \citep[][]{1943_Chandrasekhar}, which slows down their motion around the cluster centre and decreases the radius at which they reach their first apocenters. This effect on the orbit could potentially bias the measurements of the splashback radius and has been seen in simulations \citep[][]{2016ApJ...825...39M, 2022_oneil} and with observational claims \citep[cf. \citealt{2016ApJ...825...39M}]{2014JCAP...11..019A}. Dynamical friction can be minimized by using the faintest of galaxies residing in low-mass subhaloes that are affected very little by dynamical friction. The Hyper Suprime Cam (HSC) survey provides photometric galaxies down to an i-band magnitude of 26 with excellent seeing conditions \citep{2018PASJ...70S...4A}. We use galaxies from the HSC S19A internal data release \citep{2022PASJ...74..247A} and cross-correlate them with X-ray galaxy clusters selected from eFEDS. We constrain the splashback radius and compare its magnitude with the commonly used spherical overdensity boundary $R_{\rm 200m}$. For halo mass estimates, we use the galaxy shape catalogue data from HSC S16A \citep[][]{2018PASJ...70S..25M} to measure the weak lensing signal around our clusters.

We describe the different data catalogs we use in Section \ref{data}, while the measurement and the modelling techniques are described in Section \ref{measurements}. In Section \ref{results} we present the main results and compare them with earlier works. We then summarize our findings in Section \ref{conclusion}. Throughout the work, we use flat $\Lambda$CDM cosmological model with matter density $\Omega_{\rm m} = 0.27$, baryon density $\Omega_{\rm b} = 0.049$, power law index of the initial power spectrum $n_{\rm s} = 0.95$, variance of density fluctuation $\sigma_8 = 0.81$, temperature of the cosmic microwave background ${\rm T_{\rm CMB}} = 2.726 \,{\rm K}$ and the Hubble parameter $h=0.7$ as our fiducial cosmological model. The symbol $r$ represents the three-dimensional while $R$ represents the projected two-dimensional radial distance from the cluster center. We use the halo mass definition of $M_{\rm 200m}$ and corresponding halo boundary $R_{\rm 200m}$ as the radius enclosing the matter density 200 times the present matter density of the universe and the $\log$ used is the logarithm at base ten.

\section{Data}
\label{data}
\subsection{X-Ray Cluster Catalogue}
\label{clust_samp}
The eROSITA \citep{2012arXiv1209.3114M, 2021A&A...647A...1P} is a seven telescope module capable of detecting X-rays onboard the Russian-German Spectrum-Roentgen-Gamma (SRG) satellite \citep{2021A&A...656A.132S} orbiting around the Lagrange point L2. It provides a field of view of $\approx 1 {\rm deg}^2$ with excellent imaging quality. The on-axis energy resolution is $\approx 18''$ at 1.48 keV and $\approx 26''$ average angular resolution over the full field of view \citep{2021A&A...647A...1P}. The eROSITA final equatorial depth survey (eFEDS) is a small field having an area of $140\,{\rm deg}^2$  with a vignetted corrected average exposure time of $\sim 1.3$ks carried out at a depth similar to the depth to be achieved after eROSITA All-Sky Survey is complete. Thus it works as a performance verification phase for the whole survey. 

The data for the eFEDS comprises of observations in the energy band 0.2-2.3 keV. The data were collected during the first quarter of Nov 2019. This data was processed using the eROSITA Standard Analysis Software System (eSASS) pipeline for source detections \citep{2022A&A...661A...1B}. The candidate source detection for the X-ray clusters provides a detection likelihood $\mathcal{L}_{\rm det}$ and an extent likelihood $\mathcal{L}_{\rm ext}$ for each source. In order to select galaxy clusters we apply quality cuts of $\mathcal{L}_{\rm det} > 5$ and  $\mathcal{L}_{\rm ext} > 6$. The values of these thresholds were calibrated using simulations which demonstrate that these cuts tend to give a secure sample with about $80$ percent genuine clusters \citep{2020OJAp....3E..13C}. The application of the quality cuts results in a sample of 542 X-ray galaxy clusters \citep[for more detail, see][]{2022A&A...661A...1B, 2022A&A...661A...2L}. A number of these galaxy clusters were optically confirmed using the Multi-Component Match filter (MCMF) algorithm \citep{2018MNRAS.474.3324K,2019MNRAS.488..739K}. The MCMF uses galaxies from both HSC S20A and the DESI Legacy Imaging Surveys \citep{2019AJ....157..168D} to provide redshifts estimate $z_{\rm cl}$, richness $\lambda_{\rm cl}$ and secondary peak contamination fraction $f_{\rm cont}$. 
The contamination fraction provides a quantitative measure for the probability of the chance superposition in the optical. As suggested in \citet{2022A&A...661A...4K}, we use a cut of $f_{\rm cont}<0.2$ which selects an X-ray cluster sample with minimal contamination as deduced from the optical confirmation. In fact due to the luminosity cut we use for selecting the X-ray clusters, a vast majority of them (98 out of 109) have a contamination in the optical confirmation at a level less than 5\% (i.e., $f_{\rm cont}<0.05$). The full description of MCMF optimal confirmation analysis for our dataset is given in \citet{2022A&A...661A...4K}.

In the current work, we use X-ray cluster data from the publicly released eFEDS catalogue version 2.1 as described in \citet{2022A&A...661A...4K}. The catalogue provides the sky position of each of the galaxy clusters in the columns {\sc RA\_CORR, DEC\_CORR}, the redshifts and its associated error in the columns {\sc Z\_BEST\_COMB, SIGMA\_Z\_BEST\_COMB}, along with the contamination fraction in the column {\sc F\_CONT\_BEST\_COMB}. Similar, we use the columns {\sc ML\_FLUX, ML\_FLUX\_ERR} for the flux and its errors in the 0.5-2.0 keV band. We select those clusters which have less than $20$ percent relative error on both the redshift and flux measurements.

\begin{figure}
    \centering
    \includegraphics[width=\columnwidth]{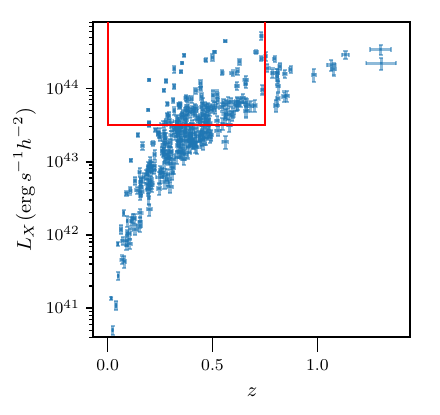}
    \caption{{\it Sample selection:} The above figure represents the scatter plot between X-ray luminosity $L_{X}$ and redshift $z$  for our eFEDS galaxy cluster sample using clusters residing in the HSC S19A footprint with $1\sigma$ errors in the x and y directions. The solid red line represents a roughly volume-limited sample selection at luminosity threshold $L_{X}>10^{43.5} {\rm erg \,s^{-1} h^{-2}}$ within the redshift of $z<0.75$ and providing 109 X-ray galaxy clusters for our analysis (without the k-correction). The eROSITA eFEDS field has roughly $90\,{\rm deg}^2$ masked overlapping region with HSC S19A.}
    \label{sel_smp}
\end{figure}

Finally, we also convert the flux values to luminosities $L_{\rm X} = 4\pi d^2_{\rm l}(z) \, f_{\rm X}$, where $d_{\rm l} (z)$ is the luminosity distance at redshift $z$ for our fiducial cosmological model. We select galaxy clusters that satisfy a threshold of $L_{\rm X}>10^{43.5} {\rm erg \,s^{-1} }h^{-2}$ and have redshift $z<0.75$ as shown in Figure \ref{sel_smp}. The solid red line shows our selection which corresponds to 109 X-ray galaxy clusters that we use for our analysis. These selection cuts help us to prune out less massive clusters at low redshift while giving us a sample that looks approximately volume limited. We note that we have not applied any k and $n_{\rm H}$ corrections to keep our sample selection simple, where $n_{\rm H}$ is the galactic neutral hydrogen column density. However, we have also checked by using the corrected values for luminosities given by \citet{2022A&A...661A...2L} and our sample changes by at most 10 percent. We found no significant effect of 10 percent change in our final constraints on the splashback feature given the uncertainties.

We will also use a subsample of 45 X-ray clusters that reside within the footprint that encompasses the HSC S16a internal data release. This subsample will be used to carry out a weak lensing analysis with the help of the first year shape catalog from the HSC survey. In order to explore any systematic effects in the weak lensing signal measurement, we use 100 times more random points in the same field constructed using the sensitivity map from the eFEDS region. We assign redshifts to these randoms by drawing with replacement from the parent redshift distribution of our cluster sample. In the absence of systematics, we do not expect any weak lensing signal to be detected around the random points. In addition, we also expect that the number of source galaxies per lens to be the same when comparing the clusters and the random points, if the sources are not physically associated with the cluster.

\subsection{Galaxy Catalogues}
We use a galaxy catalogue from the Subaru Hyper Suprime-Cam (HSC) optical imaging survey carried out using the wide-field ($1.77 \,{\rm deg}^2$)  Hyper Suprime-Cam instrument  \citep[][]{2018PASJ...70S...2K,2018PASJ...70S...1M}. The Subaru is a 8 meter class telescope situated at the summit of Maunakea in Hawaii where the median seeing is close to $\sim 0.6"$. The HSC survey collaboration has concluded a wide angle imaging survey of 1100 sq.~deg.~in the $ugrizy$ bands at a depth of $i=26$ under the Subaru Strategic Program \citep[SSP,][]{2018PASJ...70S...4A}.

We use the optical galaxy catalogue selected from the internal data release S19A for the cluster-galaxy cross-correlation and use the publicly available shape catalogue from the internal data release S16A for the weak gravitational lensing measurement. We describe the details of each catalogue in the subsections below.

\subsubsection{HSC S19A optical galaxy catalogue}
We use galaxies from an area $\sim 90\,{\rm deg}^2$ in the GAMA09H field from the internal data release S19A of the HSC survey. Following \citet[][]{2020PASJ...72...64M}, we use fluxes from the forced photometry table, and use objects brighter than z-band cmodel magnitude (galactic extinction corrected) $m_z < 24.5$. We select only those objects that have flux errors less than $20$ percent and restrict ourselves to sources that are extended in the z-band (\texttt{z\_extendedness\_value} != 0).  We note that given the redshifts of our clusters, the the $4000$\AA\, break never enters the $z$ band, which avoids any non-uniformities in the sample selection. 

In order to avoid selecting galaxies from regions which do not have sufficient depth, we use galaxies which have been observed in a minimum number of exposures in the HSC survey. Such a selection is made by using the columns \texttt{\text{[gr]\_inputcount\_value}$\geq$2} and \texttt{\text{[izy]\_inputcount\_value}$\geq$4}. We also apply various flags to remove galaxies affected by bad pixels and remove duplicates due to overlaps in the area processed by the software pipeline. In particular we use
\texttt{
\begin{itemize}
    \item z\_deblend\_skipped = False
    \item z\_cmodel\_flag\_badcentroid = False
    \item z\_cmodel\_flag = False
    \item z\_pixelflags\_edge = False
    \item z\_pixelflags\_interpolatedcenter = False
    \item z\_pixelflags\_saturatedcenter = False
    \item z\_pixelflags\_crcenter = False
    \item z\_pixelflags\_bad = False
    \item z\_pixelflags\_suspectcenter = False
    \item z\_pixelflags\_clipped = False
    \item z\_detect\_isprimary = True 
    \item z\_sdsscentroid\_flag = False
\end{itemize}
}

In addition we also use flags from the mask table in order to reduce the impact of bright stars in the HSC survey - the ghost mask, the halo mask and the blooming mask as described in \citet{2022PASJ...74..247A}. These conditions can be summarized as:
\texttt{\begin{itemize}
    \item z\_mask\_brightstar\_ghost = False
    \item z\_mask\_brightstar\_blooming = False
    \item z\_mask\_brightstar\_halo = False
\end{itemize}}

\subsubsection{HSC S16A shape catalogue}
We also infer the halo masses for our cluster sample using weak gravitational lensing. For this purpose, we use the shape information of galaxies from the incremental data release S16A that includes the first year shape catalogue \citep[][]{2019PASJ...71..114A}. This data release has slightly more data compared to the first public data release \citep[][]{2018PASJ...70S...8A} of the HSC survey. The entire S16A shape catalog spans six different fields - HECTOMAP, GAMA09H, WIDE12H, GAMA15H, XMM, and VVDS covering a total area of $136.9 \,{\rm deg}^2$ and has an effective galaxy number density of $21.5\,{\rm arcmin}^{-2}$ with a median redshift of $0.8$. 

We use data from the GAMA09H field as it overlaps with the eFEDS cluster catalogue in terms of sky area. The shape catalogue provides the sky positions for galaxies, their estimated shapes and their corresponding calibrations. The shapes of galaxies are represented by two ellipticities $(e_1, e_2) = (e\cos2\phi,e\sin2\phi)$ with $e=(a^2 - b^2)/(a^2 + b^2$), where a and b are the semimajor and semiminor axis of the galaxies \citep[][]{2002AJ....123..583B} with $\phi$ as the position angle of major axis in equatorial coordinate system. These shapes are estimated using the re-Gaussianization technique \citep{2003MNRAS.343..459H}, which has been extensively studied using data from the SDSS survey \citep{2005MNRAS.361.1287M,2012MNRAS.425.2610R,2013MNRAS.432.1544M}. 

The shape catalog also provides additive bias $(c_1 ,c_2)$, multiplicative bias $m$ corrections, the rms value for intrinsic ellipticities $e_{\rm rms}$ and the measurement error $\sigma_e$ for each galaxy. This additional calibration data were obtained by image simulations performed using an open-source software package - \textsc{GALSIM} \citep[][]{2015A&C....10..121R} that mimics observation conditions for the HSC survey \citep{2018MNRAS.481.3170M}. While computing the weak lensing signal, we assign weights $w_s$ to each galaxies using their $e_{\rm rms}$ and $\sigma_e$ as $w_s = (e^2_{\rm rms} + \sigma_e^2)^{-1}$ as is described in more detail in \citet{2018MNRAS.481.3170M}. The shape catalogue already includes various quality cuts, primary amongst which is an i-band magnitude limit of $i<24.5$ as suggested in \citet{2018PASJ...70S..25M} for studies related to weak lensing cosmology. The HSC survey provides full probability distribution function $p(z)$ for each galaxy in the shape catalogue for each of the six different methods \citep[][]{2018PASJ...70S...9T}. We use the $p(z)$ computed using the classical template fitting code - \textsc{Mizuki} \citep[][]{2015ApJ...801...20T} for the weak lensing signal measurements along with a selection cut on the $p(z)$ to extract a secured sample of background galaxies as described in Section \ref{sec-wl}. We also checked our results using other photometric redshift estimation methods and found no significant change in the inferred values for the model parameters. 

\section{Measurements and Modelling}
\label{measurements}
Here we will describe the methods used for measurements and modelling of the weak lensing profiles and the cluster galaxy cross-correlations for our sample.
\subsection{Weak Lensing Profile}
\label{sec-wl}
The weak gravitational lensing imprints a coherent tangential distortion pattern in the shapes of background source galaxies due to matter present in and around foreground lens galaxies \citep[see][for a recent review]{2015_Kilbinger, 2018ARA&A..56..393M}. These distortions can be measured as a function of projected cluster-centric comoving radial distance $R$ and are related with the projected matter distribution in the foreground lens such that
\begin{equation}
    \bar{\Sigma}(R) - \langle \Sigma (R) \rangle = \Sigma_{\rm crit} \gamma_t
    \label{esdeqn.}
\end{equation}
where the $\bar{\Sigma}(R) - \langle \Sigma (R) \rangle = \Delta \Sigma(R)$ is known as excess surface density (ESD). The quantity $\bar{\Sigma}(R) = \int_0^R \Sigma(R')  2 \pi R' \, dR' / \pi R^2$ is the average surface matter density within a projected distance $R$ while $\langle \Sigma (R) \rangle$  is the azimuthally averaged surface matter density at the same distance. The quantity $\gamma_t$ denotes the average tangential shear, and $\Sigma_{\rm crit}$ denotes the critical surface density, a geometric factor that quantifies the lensing efficiency of a given lens-source pair, such that,
\begin{equation}
    \Sigma_{\rm crit} = \frac{c^2}{4\pi G} \frac{D_{\rm a} (z_{\rm s})}{(1 + z_{\rm l})^2 D_{\rm a}(z_{\rm l}) D_{\rm a}(z_{\rm l}, z_{\rm s})}\,.
\end{equation}
Here the quantities $D_{\rm a} (z_{\rm s}), D_{\rm a}(z_{\rm l})$ and $D_{\rm a}(z_{\rm l}, z_{\rm s})$ are the angular diameter distances between us and the source at redshift $z_{\rm s}$, us and the lens at redshift $z_{\rm l}$ and between a given lens-source pair, respectively. The $(1+z_{\rm l})^2$ factor in the denominator is related to our use of comoving coordinates \citep[][]{2006MNRAS.372..758M}.

Following \citet[][]{2018PASJ...70S..25M}, we compute the ESD $\Delta \Sigma(R_i)$ for our lensing sample using the HSC background source galaxies at ten logarithmic cluster centric projected comoving radial distance bins $R_i$ ranging from 0.2 to 2 ${\rm h^{-1} Mpc}$, such that
\begin{multline}
    \Delta \Sigma(R_i) = \frac{1}{1 + m} \left( \frac{\sum_{{\rm ls}\in R_i} w_{\rm ls} e_{t,{\rm ls}} \langle \Sigma_{\rm crit}^{-1} \rangle^{-1}}{2\mathcal{R} \sum_{{\rm ls}\in R_i} w_{\rm ls}} \right)\\ - \frac{1}{1 + m} \left(\frac{\sum_{{\rm ls}\in R_i} w_{\rm ls} c_{t,{\rm ls}} \langle \Sigma_{\rm crit}^{-1} \rangle^{-1}}{\sum_{{\rm ls}\in R_i} w_{\rm ls}} \right)\,.\label{eq:dsig}
\end{multline}
Here $e_{t,{\rm ls}}$ and $c_{t,{\rm ls}}$ are the components of the ellipticities $(e_1, e_2)$ and additive bias $(c_1 , c_2)$ in the tangential direction to the line joining the lens and the source. The summation is over all the lens-source pair ${\rm ls}$ having separation of $i^{\rm th}$ radial bin $R_i$. We also use minimum variance weights such that $w_{\rm ls} = w_{\rm s} \langle \Sigma_{\rm crit}^{-1} \rangle^2 $. The quantity $\langle \Sigma_{\rm crit}^{-1} \rangle$ is the average of the inverse surface critical density over the probability distribution of source redshift $p(z_{\rm s})$ given by
\begin{equation}
    \langle \Sigma_{\rm crit}^{-1} \rangle = \frac{4\pi G (1+z_{\rm l})^2}{c^2} \int_{z_{\rm l}}^\infty \frac{D_{\rm a} (z_{\rm l}) D_{\rm a} (z_{\rm l},z_{\rm s})}{ D_{\rm a} (z_{\rm s})} p(z_{\rm s}) dz_{\rm s}\,.
\end{equation}

The $1+m$ factor in the denominator of Eq.~\ref{eq:dsig} corresponds to the multiplicative bias with $m = \sum_{{\rm ls} \in R_i} w_{\rm ls} m_{\rm s} / \sum_{{\rm ls} \in R_i} w_{\rm ls}$ and $\mathcal{R}$ is the shear responsitivity which corrects for the response of ellipticities to an applied value of shear and is computed using $e_{\rm rms}$ \citep{2002AJ....123..583B} as, 
\begin{equation}
    \mathcal{R} = 1 - \frac{\sum_{{\rm ls} \in R_i} w_{\rm ls} e^2_{{\rm rms},{\rm ls}}}{\sum_{{\rm ls} \in R_i} w_{\rm ls}}\,.
\end{equation}

In order to reduce contamination of the lensing signal by foreground galaxies or galaxies correlated with our clusters, we only use sources which satisfy
\begin{equation}
\int_{z_{\rm max} + z_{\rm diff}}^\infty  p(z_{\rm s}) dz_{\rm s} > 0.99 \,.
\end{equation}
We use $z_{\rm max} = 0.75$ as the maximum redshift of our lensing sample and $z_{\rm diff} = 0.2$ as an additional offset for better selection of the background. We also use galaxies with a photo-z quality cut of $\texttt{photo\_z\_risk\_best\_value} < 0.5$. Further, we also apply a multiplicative bias $m_{\rm sel}$ related to the selection of galaxies above the resolution threshold ($R_2 \geq 0.3$) which is used to select source galaxies during the construction of the shape catalog \citep[see ][]{2018MNRAS.481.3170M}. This selection bias is related to the probability density $p(R_2)$ at the threshold $R_2 = 0.3$ such that $m_{\rm sel} = A p(R_2 =0.3)$ with $A=0.00865$. The probability $p(R_2=0.3)$ is computed using the lens-source weights $w_{\rm ls}$ for individual radial bin $R_i$.

We apply the redshift selection cuts on source galaxies to select the background galaxy population. Any residual galaxies left in our source population that are associated with the cluster could potentially dilute the weak lensing signal. We can correct for such a dilution by multiplying the signal by a boost factor $C(R_i)$ \citep[for example][]{2004MNRAS.353..529H, 2005MNRAS.361.1287M, 2015ApJ...806....1M} which is the ratio between weighted lens-source pair counts to the weighted random-source pair counts in a given radial bin $R_i$ such that,
\begin{equation}
    C(R_i) = \frac{N_{\rm r} \sum_{{\rm ls}\in R_i} w_{\rm ls}}{N_{\rm l} \sum_{{\rm rs}\in R_i} w_{\rm rs}}\,.
    \label{boost_eqn.}
\end{equation}
Here the quantities $N_{\rm r}$ and $N_{\rm l}$ are the number of random points and lenses in our sample, respectively. The weight $w_{\rm ls}$ and $w_{\rm rs}$ correspond to lens-source and random-source pairs. We use 100 different random realizations in our work such that they have the same number as lenses within the same sky coverage, and also satisfy the same star mask, and by construction, have the same redshift distribution as our lensing sample. We find that the boost parameters are mostly consistent with unity demonstrating the utility of our redshift cuts, so we neglect them in this study. We refer to Appendix \ref{apx_boost} for more discussion on boost parameter measurements for our lensing sample.

Along with the correction for the dilution of ESD signal using the boost parameters, we also check for any systematic bias due to the use of photometric redshifts for the source galaxies while computing the critical surface density. Following eqn. 5 in \citet{2008MNRAS.386..781M}, we compute the magnitude of such biases $b(z_l)$ for a lens at redshift $z_l$ using
\begin{equation}
    \frac{\Delta \Sigma}{\Delta \Sigma^t} = 1 + b(z_{\rm l})  = \frac{\sum_{\rm s} w_{\rm ls} \langle \Sigma^{-1}_{{\rm crit},{\rm ls}}\rangle^{-1} \left(\Sigma^{t}_{{\rm crit}, {\rm ls}}\right)^{-1}}{\sum_{\rm s} w_{\rm ls}}\,.
\end{equation}
Here the quantities with superscript $t$ represent the true values of the corresponding quantities and the sum runs over all the source galaxies. Ideally the bias needs to be estimated using spectroscopic redshifts for a representative subsample of source galaxies.  However, given the depth of the HSC survey, a reasonably large survey with such spectroscopic redshift is not available. Therefore, we use robust estimates of photometric redshifts from the COSMOS-30 band photometry \citep[][]{2009ApJ...690.1236I} and assume it to be a much more realistic estimate of the redshifts of our source galaxies. We also include weights for each source galaxy $w_{\rm som}$ that match the colour and magnitudes distribution of COSMOS-photoz galaxies to our source galaxy sample. These $w_{\rm som}$ are included in the $w_{\rm ls}$ while doing the computations as done in previous studies \citep[for e.g.][]{2012MNRAS.420.3240N, 2019ApJ...875...63M, 2019PASJ...71..107M}. We then use eqn. 23 from \citet{2012MNRAS.420.3240N} to assign appropriate weights for our lenses to compute the average bias for the ESD signal. For our sample, we obtain three percent bias on average, which is negligible given the statistical uncertainty in the signal measurements.

We also subtract the ESD signal around random points from the signal computed from our lensing sample to correct for any scale dependent systematics \citep[for more details refer to][]{2004AJ....127.2544S, 2005MNRAS.361.1287M, 2017MNRAS.471.3827S}. We compute the random signal by averaging over ESD measurements from 100 different realizations of randoms.

The presence of the same source galaxy at different radial bins for different galaxy clusters can give rise to a covariance between the ESD measurements in the different radial bins. We quantify this covariance by randomly rotating the shapes of our source galaxies which washes away the shear signal imparted on them and allows us to estimate the covariance due to shape noise. In our study, we utilize 200 different random rotations and use the ESD measurements in each case to estimate the covariance, as discussed in Appendix \ref{meascov}. As we model signal up to $1 \, h^{-1}{\rm  Mpc }$ we expect shapenoise to dominate over large scale structure contribution on these length scales. So, we use the shape noise covariance for our analysis and infer the halo masses for our lensing sample.

We model the dark matter distribution around clusters as a NFW density profile $\rho_{\rm nfw} (r)$ at a three-dimensional radius $r$,

\begin{align}
    \rho_{\rm nfw}(r) &= \frac{\delta_{\rm c} \rho_{\rm m}}{\left(\frac{r}{r_{\rm s}}\right)\left(1 + \frac{r}{r_{\rm s}}\right)^2}\,, \\
    \delta_{\rm c} &= \frac{200}{3} \frac{c^3}{\ln(1+c) - c/(1+c)}\,.
\end{align}
Here $r_{\rm s} = r_{\rm 200m} / c$ is the scale radius of the halo with $r_{\rm 200m}$ as the three dimensional radius of a sphere enclosing the average density of $200$ times the present matter density $\rho_{\rm m}$ of the Universe,
\begin{equation}
r_{\rm 200m} = \left(\frac{3 M_{\rm 200m}}{4\pi 200 \rho_{\rm m}}\right)^{1/3}\,.    
\end{equation}
So given the model we just need two parameters $\Theta = (M_{\rm 200m}, c)$ the halo mass $M_{\rm 200m}$ and the concentration parameter $c$ to predict the ESD profile $\Delta \Sigma(R)$. We use uninformative flat priors for both $\log M_{\rm 200m}$ and  $c$ given in Table \ref{paramprior}, which are conservative given the expectation of concentration for cluster scale dark matter halos \citep[for e.g.][]{2007_Comerford}. In practice, we use the analytical form for $\Delta \Sigma(R)$ given by eqn. 14 in \citet{1999_wright} for the case of the NFW profile.  The weak lensing signal is thus given by
\begin{align}
\Delta\Sigma(R) &= {r_{\rm s}\delta_{\rm c} \rho_{\rm m}}\ f(R/r_{\rm s})\,, \\
f(x) &= 
\begin{cases}
g_{<}(x) &  $x<1$\,, \\\\
\frac{10}{3} + 4\ln\left(\frac{1}{2}\right) &  $x=1$ \,,\\\\
 g_{>}(x) &  $x>1$\,, \\\\
\end{cases}
\end{align}
where functions $g_{<}(x)$ and $g_{>}(x)$ are calculated as
\begin{multline}
g_{<}(x) = \frac{8\,{\rm arctanh} \sqrt{(1-x)/(1+x)}}{x^2\sqrt{1-x^2}} + \frac{4}{x^2}\ln\left(\frac{x}{2}\right)\\ -\frac{2}{x^2 -1} + \frac{4\,{\rm arctanh}\sqrt{\left(1-x\right)/\left(1+x\right)}}{\left(x^2 - 1\right)\left(1 -x^2\right)^{1/2}}\,,
\end{multline}

\begin{multline}
g_{>}(x) = \frac{8\,{\rm arctanh} \sqrt{(x-1)/(1+x)}}{x^2\sqrt{x^2 - 1}} + \frac{4}{x^2}\ln\left(\frac{x}{2}\right)\\ -\frac{2}{x^2 -1} + \frac{4\,{\rm arctanh}\sqrt{\left(x-1\right)/\left(1+x\right)}}{\left(x^2 -1\right)^{3/2}}\,.
\end{multline}

In principle, previous works have also added a point mass contribution for the baryonic component of the central galaxy of the cluster \citep[e.g., ][]{2015MNRAS.449.2128K}. However for the scales of our interest, the dark matter contribution is the dominant component. We also do not split the dark matter contribution into 1-halo and 2-halo terms \citep[for more details, see][]{2013MNRAS.435.2345H, 2016PhRvL.116d1301M} or use an off-centering kernel \citep{2007arXiv0709.1159J} in order to account for the possible misidentification of the cluster centre. Given the scales we fit and the statistical errors on our mass estimates, none of these effects would cause a significant difference to our conclusions. Furthermore we also do not carry out a halo occupation distribution (HOD) based modelling \citep[e.g.,][]{2000MNRAS.318..203S, 2002PhR...372....1C, 2013MNRAS.430..725V} as the quantity of our interest is the average halo mass rather than the entire distribution of the halo masses. At the high mass end corresponding to our galaxy clusters, the distribution is expected to be peaked due to the presence of the exponential tail of the mass function. Due to all these reason we limit our modelling to the simple NFW profile based modelling scheme as described above. In future, we expect the splashback radius measurements using X-ray clusters to improve further, which will require a more detailed analysis of the weak lensing signal of these galaxy clusters.

\subsection{Galaxy Number Density}
\label{sec-c-g-xcorr}
We follow the method used in \citet{2016ApJ...825...39M} for the measurement of the cluster galaxy cross-correlation measurements. We use the Davis-Peebles estimator \citep{1983ApJ...267..465D} to compute the projected cross-correlation between our X-ray clusters with the HSC photometric galaxies. The projected cross-correlation $\xi_{\rm 2D}(R)$ at a projected comoving radius $R$ is given by
\begin{equation}
    \xi_{\rm 2D}(R) = \frac{D_{\rm c} D_{\rm g}}{R_{\rm c} D_{\rm g}} - 1 \,,
\end{equation}
where $D_{\rm c} D_{\rm g}$ are the pair counts between clusters-galaxies and $R_{\rm c} D_{\rm g}$ are the pair counts between cluster randoms-galaxy at a comoving projected separation of $R$. The number counts for the random points  are normalized to account for the difference in the number of clusters and that of randoms. We measure the signal over nine logarithmically spaced comoving projected radial bins from 0.1 to 2.8 $ h^{-1}{\rm Mpc}$. We use 40 times more randoms than the number of galaxy clusters in our sample.

The flux limit of the photometric survey can affect the galaxy distribution around the galaxy clusters by detecting many fainter galaxies around the clusters closer to us and can bias our measurements. To remove such biases, we use the photometric galaxies having z-band absolute magnitude cut $M_{z} -5\log h < -19.36$, which corresponds to an apparent magnitude limit of  $m_z = 24.0$ at redshift $z = 0.75$ which ensures the population of galaxies has similar intrinsic properties around each cluster in our sample even though they are at different redshifts. We estimate the covariance of our measurements using 20 jackknife \citep{miller_1974} regions that have an approximately square shape and an area of $5\,{\rm deg^2}$. The side of the squares corresponds to $47 \, h^{-1}{\rm Mpc}$ at the median redshift $z=0.46$ for our sample and is larger than the radial range for signal measurement, justifying our choice of jackknife region size. We have shown the corresponding covariance in Appendix \ref{meascov}.

We parameterize the density profile with the functional form suggested by \citet[][]{2014ApJ...789....1D} in order to model the cross correlation signal as is standard practice in the literature \citep[][]{2016ApJ...825...39M, 2017ApJ...841...18B,2018ApJ...864...83C, 2018PASJ...70S..24N, 2019ApJ...874..184Z, 2019MNRAS.487.2900S, 2020PASJ...72...64M,2021MNRAS.507.5758S}. The density profile $\xi_{\rm 3D}(r)$ consists of an inner Einasto profile \citep{1965TrAlm...5...87E} $\rho_{\rm in}(r)$ and an outer powerlaw profile $\rho_{\rm out}(r)$ connected by a smooth transistion function $f_{\rm trans} (r)$ at a three dimensional radial distance of $r$,
\begin{align}
    \xi_{\rm 3D}(r) &= \rho_{\rm in}(r) f_{\rm trans} (r) + \rho_{\rm out}(r)\,,\\
    \label{dk14_in}
    \rho_{\rm in} (r) &= \rho_{\rm s}\exp\left( -\frac{2}{\alpha}\left[\left( \frac{r}{r_{\rm s}}\right)^\alpha - 1 \right]\right)\,,\\
    \label{dk14_out}
    \rho_{\rm out}(r) &= \rho_{\rm o} \left(\frac{r}{r_{\rm out}}\right)^{- s_{\rm e}}\,,\\
    \label{dk14_trans}
    f_{\rm trans}(r) &= \left(1 + \left(\frac{r}{r_{\rm  t}}\right)^\beta\right)^{-\gamma/\beta}\,.
\end{align}
We compute the two-dimensional cross-correlation profile $\xi_{\rm 2D}(R)$ at the projected radius $R$ by integrating the three-dimensional profile $\xi_{\rm 3D}(r)$ along the line of sight direction $\pi$,
\begin{equation}
\label{xi2d}
\xi_{\rm 2D}(R) = \frac{1}{R_{\rm max}} \int_0^{R_{\rm max}} \xi_{\rm 3D}\left(\sqrt{(R^2 + \pi^2)}\right) d\pi\,.
\end{equation}
Following \citet{2016ApJ...825...39M} we fix the maximum projected length of $R_{\rm max} = 40 \,h^{-1} {\rm Mpc}$. Our modelling scheme eqn. \ref{xi2d}  comprises of nine parameters - $\rho_{\rm s}, \alpha, r_{\rm s}, \rho_{\rm o}, r_{\rm out}, s_{\rm e}, \beta, r_{\rm t}, \gamma$. As $\rho_{\rm o}$ and $r_{\rm out}$ are degenerate with each other, we fix $r_{\rm out} = 1.5\,{ h^{-1} {\rm Mpc}}$ which gives us a total of eight parameters $\Theta = (\rho_{\rm s}, \alpha, r_{\rm s}, \rho_{\rm o}, s_{\rm e}, \beta, r_{\rm t}, \gamma)$ modelling scheme.

\subsection{Model Fitting}
We use the Bayesian analysis to get the posterior probability for our model parameters $\Theta$ given the data $\mathcal{D}$ with priors $P(\theta)$ on the parameters as given in Table \ref{paramprior}. In our work, we perform separate fits to the weak lensing and projected galaxy number density measurements. We use a flat prior on most parameters for both models except for Gaussian priors on $\log \alpha$, $\log \beta$ and $\log \gamma$. The Gaussian priors for $\log \alpha$, $\log \beta$ and $\log \gamma$ are similar to those commonly used in literature \citep[for eg., ][]{2016ApJ...825...39M,2019MNRAS.487.2900S,2020PASJ...72...64M} for splashback radius studies as they are motivated from simulations \citep[][]{2008MNRAS.387..536G,2014ApJ...789....1D} and are characteristic of typical cluster scale halos.
 
From the Bayes theorem, we can write the posterior $P(\Theta|\mathcal{D})$ as 
\begin{equation}
    P(\Theta|\mathcal{D}) \propto P(\mathcal{D}|\Theta) P(\Theta) \,,
\end{equation}
where $P(\mathcal{D}|\Theta)$ is the likelihood of the data given model parameters, and we are using a Gaussian likelihood given by 
\begin{align}
    P(\mathcal{D}|\Theta) \propto \exp\left( -\frac{\chi^2 (\Theta)}{2}\right) 
\end{align}
with $\chi^2(\Theta) = [\mathcal{D} - \mathcal{M}(\Theta)]^T C^{-1} [\mathcal{D} -\mathcal{M}(\Theta)]$, $\mathcal{D}$ is the data vector with $\mathcal{M}(\Theta)$ as the model prediction vector given the parameter $\Theta$. The noise in the covariance matrix can result in a bias when inverted to obtain the $\chi^2$. We follow eqn. 17 in \citet[][]{2007A&A...464..399H} in order to account for this bias. We use the affine invariant Monte Carlo Markov Chain \citep[MCMC][]{2010CAMCS...5...65G} sampler as provided by the python package \texttt{emcee} \citep{2013PASP..125..306F} to infer the posterior distribution for our model parameters given the measurements.  In our work, we run separate MCMCs to infer parameter posteriors for the weak lensing and cross-correlation measurements for our cluster sample. We use 256 walkers with 7500 steps, and the the chains converged within 1000 steps, and the posterior distribution of the halo mass for the weak lensing fits and the 3-d splashback radius for the cross-correlation fits did not show any significant shifts.

\begin{table}
    \centering
    \begin{tabular}{|p{2.5cm}||p{2.5cm}|}
    \hline
    \multicolumn{2}{|c|}{Model Parameters: galaxy number density}\\
    \hline
    Parameter & Prior \\
    \hline
     ${\log \rho_{\rm s}}$ & Flat[-3, 5] \\
	 ${\log \alpha}$ &  Gauss($\log(0.2)$, 0.6)\\
	 ${\log r_{\rm s}}$ & Flat[$\log (0.1), \log (5.0)$]\\
	 ${\log \rho_{\rm o}}$ & Flat[-1.5, 1.5]\\
	 $s_e$ & Flat[0.1, 4]\\
	 ${\log r_{\rm t}}$ & Flat[$\log (0.5), \log (1.6)$]\\
	 ${\log \beta}$ & Gauss($\log(6.0)$, 0.2) \\
	 ${\log \gamma}$ & Gauss($\log(4.0)$, 0.2) \\
	 \hline
	 \multicolumn{2}{|c|}{Model Parameters: weak lensing profile}\\
	 \hline
     Parameter & Prior \\
     \hline
	 $\log[{M_{\rm 200m} / h^{-1}{\rm  M_\odot}}]$ & Flat[12, 16] \\
	 $c$ & Flat[0, 20]\\
    \hline
    \end{tabular}
    \caption{{\it Parameter priors:} The table provides the prior distributions used for running the MCMC chains on our model parameters. 
    The Flat[a, b] denotes uniform prior in the interval (a, b), and Gauss($\mu$,$\sigma$) is the Gaussian priors with $\mu$ mean and $\sigma$ standard deviation. The parameters $\log \rho_{\rm s},\,\log \alpha,\,\log r_{\rm s}$ models the inner Einasto profile with $\rho_{\rm in}(r)$; $s_e,\,\log \rho_{\rm o}$ for outer profile $\rho_{\rm out}(r)$ and $\log r_{\rm t},\,\log \beta,\, \log \gamma$ for the transition function $f_{\rm trans}(r)$ and are used to model the galaxy number density profile. The halo mass $\log[{M_{\rm 200m} / h^{-1} {\rm  M_\odot}}]$ and concentration $c$ are for the NFW profile-based modelling of the weak lensing measurements for halo mass calibrations.}
    \label{paramprior}
\end{table}

\section{Results and Discussions}
\label{results}

In this section, we describe our results for the weak lensing and projected cluster-galaxy cross-correlation on the X-ray cluster sample. The weak lensing measurements allows us to infer the average halo mass of our sample, while the cross-correlation analysis will allow us to infer the location of the splashback radius in three dimensions. We will combine the two to present constraints on $r_{\rm sp}/r_{\rm 200m}$.

\subsection{Halo Mass for Galaxy Clusters}
\label{meas-halo}
We follow the methodology described in Sec \ref{sec-wl} and measure the weak lensing signal $\Delta \Sigma(R)$ around our X-ray clusters in ten projected logarithmically-spaced radial bins from the X-ray cluster centre with comoving distances in the range of $[0.1,
2.0]\,h^{-1} {\rm  Mpc}$. We then model the signals using a simple NFW profile  and infer the corresponding halo mass $M_{\rm 200m}$ and use it to obtain the spherical overdensity size estimate $r_{\rm 200m}$. 

\begin{figure}
    \centering
    \includegraphics[width=\columnwidth]{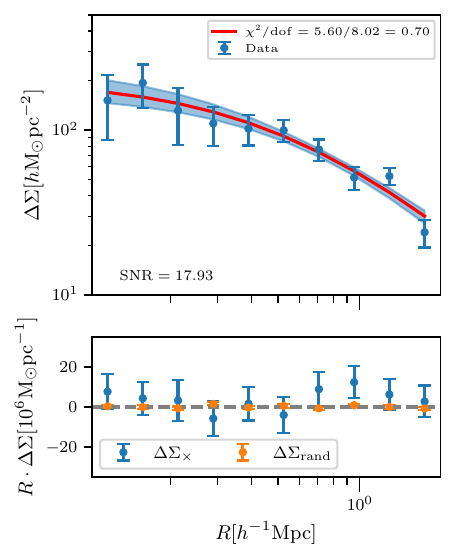}
    \caption{{\it Weak lensing signals:} In the top panel, the blue data points correspond to our weak lensing measurements with shape noise errors for a nearly volume-limited selected X-ray cluster sample (see Figure \ref{sel_smp}) with signal-to-noise given in the bottom right corner. The solid red line denotes the best fit model prediction with $\chi_{\rm red}^2 \equiv \chi^2 / {\rm dof}$ value given in the upper right corner. The blue-shaded region represents the 68 percentile around the median model predictions, and Table \ref{modparam} presents the constrained model parameter values. The bottom panel shows the systematic checks for our weak lensing signal $\Delta \Sigma$ measurements. $\Delta \Sigma_{\times}$ denotes the cross component of weak lensing signal with shape noise errors, and $\Delta \Sigma_{\rm rand}$ shows the measurements around the random points with errors over mean computed using the random realisations. The grey dashed zero line shows the expected null signal.}
    \label{esd_sig}
\end{figure}

In the top panel of Figure \ref{esd_sig}, the blue data points represent our measurements $\Delta \Sigma$ and the errors are a result of shape noise. Our measurements have a total signal-to-noise of $17.93$. The weak lensing signal shows the expected behavior and decreases as a function of the projected radial distance to the cluster centre. The best fit model is shown by the solid red line and corresponds to a reduced chi-squared $\chi^2_{\rm red} = 5.6/8.02$, where the degrees of freedom have been computed using eq.~29 in  \citet{2019PhRvD..99d3506R}.  The blue shaded region indicates the 68 percent of the model predictions with $\chi^2$ closest to the best fit. It shows that our model is a good description of our measurements. 

The bottom panel of Figure \ref{esd_sig}  shows the results of systematic tests for our signal measurements. The blue points with errors show the cross-component $\Delta \Sigma_{\times}$ of the signal. This signal is consistent with zero shown as the dashed grey line with a $\chi^2$ of 5.26 for 10 degrees of freedom and a corresponding p-value of 0.86. The weak gravitational lensing is a result of the correlated dark matter distribution around our sample of lenses. Therefore, we expect a null signal when we measure the same around random points albeit any systematics. Similar to the $\Delta\Sigma_{\times}$, we also obtain a null signal $\Delta\Sigma_{\rm rand}$ for the random points residing in the survey region where the error is computed as the scatter from 100 random realizations. These two systematic checks show that our signal measurements are robust.

\begin{figure}
    \centering
    \includegraphics[width=\columnwidth]{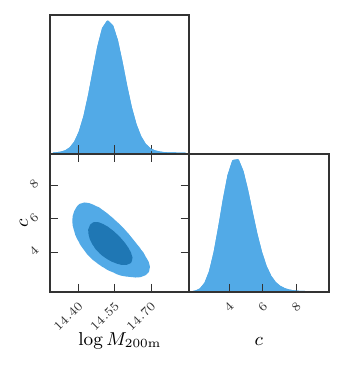}
    \caption{{\it Mass profile fits posteriors:} The above contour plot shows the degeneracy between our cluster halo mass $\log M_{\rm 200m}$ and concentration parameter $c$, respectively. The light and dark blue shaded contours are 68 and 95 confidence levels.}
    \label{nfw_post}
\end{figure}

In Figure \ref{nfw_post}, the light and dark shaded blue regions show the 68 and 95 percent credible regions in the inference of the halo mass and its concentration parameter. We obtain 13 percent constraints on the halo mass, i.e., $\log[{M_{\rm 200m}/ h^{-1}{\rm M_\odot}}] = 14.52 \pm 0.06$ and approximately $20$ percent on the concentration parameter $c = 4.49_{-0.79}^{+0.90}$. The $\Lambda{\rm CDM}$ expectation for the concentration at the halo mass of our interest computed using the concentration-mass relation by \citet{2019ApJ...871..168D} from the package {\sc COLOSSUS} \citep{2018ApJS..239...35D} is equal to $4.83$ which is consistent with our inference. We have also cross-checked our measured halo masses with those made using a joint calibration of weak lensing and cluster abundance for our X-ray cluster sample \citep{2022_Chiu_b}\footnote{This study uses $M_{\rm 500c}$ as the halo mass definitions and we use an concentration-mass relation by \citet{2019ApJ...871..168D} in {\sc COLOSSUS} to convert the masses into the definition used in our work $M_{\rm 200m}$.} and found $\log[{M_{\rm 200m}/ h^{-1} {\rm M_\odot}}] = 14.58$ , which is consistent with our measurements. In principle, the weak lensing measured halo mass may be different from the true halo mass due to various projection effects as well as triaxiality of the halo \citep[see e.g.,][]{2011_Becker}. In the case of HSC WL of eFEDS clusters, this bias is consistent with 0, and is known to about 3-6 percent \citep{Grandis:2021,2022_Chiu_b, 2022_chiu}. It is thus negligible compared to our statistical mass uncertainty of 13 percent.

Based on our inferred halo mass ${M_{\rm 200m}}$, we derive the corresponding value of the traditional halo boundary ${r_{\rm 200m}} = 1.75 \pm 0.08 \, h^{-1} {\rm  Mpc}$. In Section \ref{meas-splash}, we will use this constraint on ${ r_{\rm 200m}}$ and compare it to our inferred value of the splashback radius.

\subsection{Splashback radius of X-ray Clusters}
\label{meas-splash}
We follow the methodology described in Section \ref{sec-c-g-xcorr} to measure the cross-correlation signal between our eFEDS X-ray cluster sample and the HSC S19A galaxies in nine logarithmically spaced comoving projected radial bins in the range $[0.1, 2.8] \, h^{-1}{\rm Mpc}$ away from the X-ray centre of our clusters. We model these measurements using Eq.~\ref{xi2d} for the projected cross-correlation profile $\xi_{\rm 2D}(R)$ and infer the radial location for the splashback feature $R_{\rm sp}$. We also inferred the corresponding three-dimensional cross-correlation profile $\xi_{\rm 3D} (r)$ and the value of the splashback radius $r_{\rm sp}$.

\begin{figure*}
    \centering 
    \includegraphics[width=\textwidth]{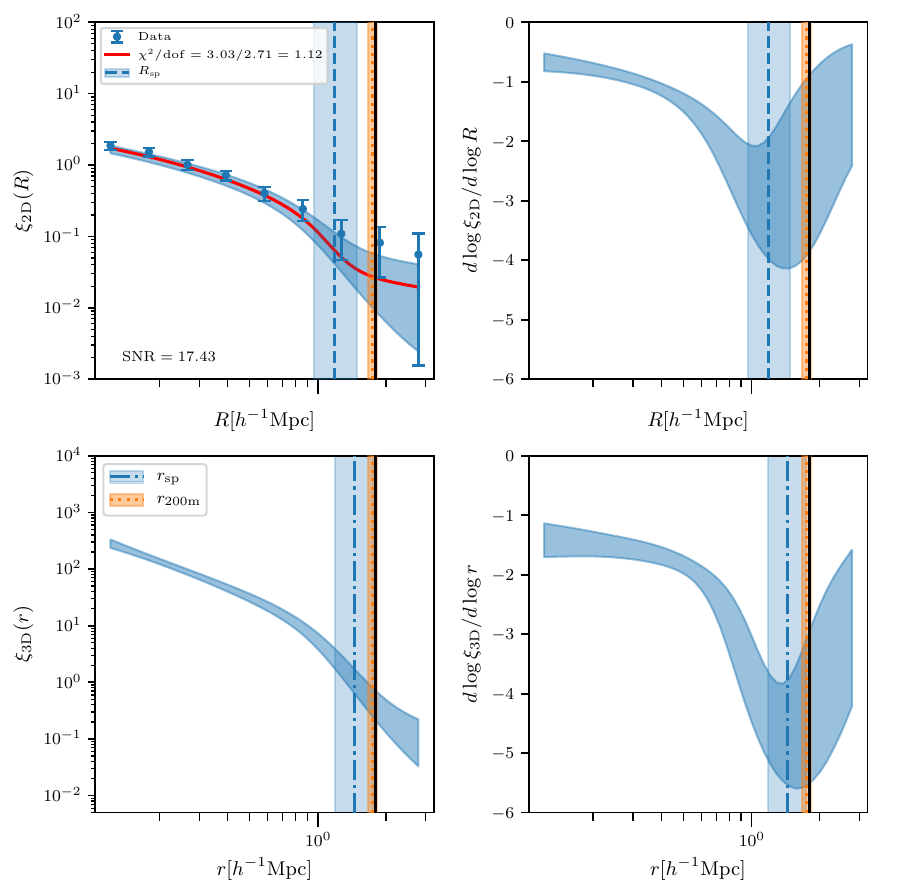}
    \caption{{\it Galaxy number density profile:} The blue data points in the top left panel show our measurements for cross-correlation $\xi_{\rm 2D}(r)$ between the eFEDS X-ray galaxy cluster and HSC S19A optical galaxies with signal-to-noise given in the bottom left corner. The solid red line represents our best fit model predictions with $\chi_{\rm red}^2 \equiv \chi^2 / {\rm dof}$ value given in the top left corner, and the dark blue shaded regions denoted the 68 percentile around the median model predictions. The top right panel shows the inferred logarithmic derivative of model predictions for projected cross-correlation profile $\xi_{\rm 2D} (r)$. In the top row, The blue dashed vertical line represents the median value for the inferred two-dimensional splashback radius $R_{\rm sp}$ with a light blue 68 percentile shaded error region around it. Similarly, the bottom row shows the corresponding three-dimensional $\xi_{\rm 3D} (r)$ profile and the associated logarithmic derivative. In the bottom row, The blue dotted dashed vertical line with a light blue shaded region around it indicates our constraints on the three-dimensional splashback radius $r_{\rm sp}$ estimates. In all the panels, the orange dotted vertical line with a shaded region shows the weak lensing calibrated three-dimensional spherical overdensity size $r_{\rm 200m}$ using the source galaxies from the HSC S16A shape catalogue. The solid black vertical line is the expectation value for $r_{\rm sp}$ from $\Lambda{\rm CDM}$ prediction \citep{2015ApJ...810...36M}. We provide the model parameter constraints with the corresponding splashback radius and spherical overdensity sizes in Table \ref{modparam}.}
    \label{c-g-xcorr}
\end{figure*}

In Figure \ref{c-g-xcorr}, the blue data points with errors correspond to our measurements of the cross-correlation signal. The errors were obtained using the jackknife technique. The cross-correlation signal  $\xi_{\rm 2D}(R)$ is detected with a signal-to-noise ratio of $17.43$. As expected the projected number of galaxies correlated with the cluster centre decrease as we move further away from the cluster centre. The solid red line shows the best fit model and corresponds to a reduced chi-square $\chi^2_{\rm red} = 1.12$ and computed using the effective degrees of freedom ${\rm dof_{eff}}=2.71$ (see eqn. 29 in \cite{2019PhRvD..99d3506R}). The  dark blue shaded regions show the 68 percent credible regions which show that the measurements are in reasonable agreement with the expectations from the model. In the top left and right panel, the blue vertical dashed line and the shaded region shows the median of the inferred location for the projected splashback radius $R_{\rm sp}$ along with the 68 percent credible interval. Furthermore, the orange dotted vertical line with the shaded region shows the inferred value for the boundary of the halo in the traditional sense - $r_{\rm 200m}$ inferred from our weak lensing analysis as described in Section \ref{meas-halo}.

\begin{table}
    \centering
    \begin{tabular}{|p{3cm}||p{3cm}|}
    \hline
    \multicolumn{2}{|c|}{Model Parameters: galaxy number density}\\
    \hline
    Parameters & Constraints \\
    \hline
     ${\log \rho_{\rm s}}^{*}$ & $0.78^{+0.55}_{-0.67} \vspace{0.1cm} $\\
	 ${ \log \alpha}$ & $-0.71^{+0.33}_{-0.30} \vspace{0.1cm} $\\
	 ${\log r_{\rm s}}^{*}$ & $0.09^{+0.40}_{-0.32} \vspace{0.1cm} $\\
	 ${\log \rho_{\rm o}}^{*}$ & $-0.85^{+0.57}_{-0.45} \vspace{0.1cm} $\\
	 ${s_{\rm e}}$ & $1.69^{+0.75}_{-0.85} \vspace{0.1cm} $\\
	 ${\log r_{\rm t}}$ & $0.00^{+0.08}_{-0.08} \vspace{0.1cm} $\\
	 ${ \log \beta}$ & $0.78^{+0.06}_{-0.06} \vspace{0.1cm} $\\
	 ${ \log \gamma}$ & $0.60^{+0.06}_{-0.06} \vspace{0.1cm} $\\
	 $R_{\rm sp} /h^{-1}{\rm  Mpc}$ & $1.19^{+0.30}_{-0.22} \vspace{0.1cm} $\\
	 $r_{\rm sp}/h^{-1}{\rm  Mpc}$ & $1.45^{+0.30}_{-0.26} \vspace{0.1cm} $\\
	 $\frac{d\log \xi_{\rm 3D}}{d\log r}|_{r = r_{\rm sp}}$ & $-5.05^{+0.88}_{-0.73} \vspace{0.1cm} $\\
	 $\chi^2_{\rm sp}/{\rm dof_{eff}}$ & 3.03/2.71 \vspace{0.1cm}\\
	 \hline
     \multicolumn{2}{|c|}{Model Parameters: weak lensing profile}\\
     \hline
     Parameters & Constraints \\
     \hline
	 $\log[{M_{\rm 200m}/ h^{-1}{\rm M_\odot}}]$ & $14.52_{-0.06}^{+0.06}\vspace{0.1cm}$ \\
	 $c$ & $4.49_{-0.79}^{+0.90} \vspace{0.1cm}$\\
	 $r_{\rm 200m}/h^{-1}{\rm  Mpc}$ & $1.75^{+0.08}_{-0.08} \vspace{0.1cm} $\\
	 $\chi^2_{\rm wl}/{\rm dof_{eff}}$ & 5.6/8.02 \vspace{0.1cm}\\
    \hline
    \end{tabular}
    \caption{{\it Parameter constraints:} The table provides the median values for the model parameter constraints with errors based on the 16 and 84 percentile of the posterior distribution obtained from modelling the measurements for our eFEDS X-ray galaxy cluster sample. The first eight rows present the parameter constraints for fitting the galaxy number density profiles. The two-dimensional $R_{\rm sp}$ and three-dimensional $r_{\rm sp}$ splashback radius estimates along with the logarithmic slope values at $r_{\rm sp}$ as given in rows 9-11. Row 12-14 presents the parameter values from the weak lensing analysis and the corresponding spherical overdensity size $r_{\rm 200m}$ value. The last two rows present the best fit $\chi^2$ values for the weak lensing $\chi^2_{\rm wl}$ and galaxy density profile $\chi^2_{\rm sp}$ with the associated effective degrees of freedom ${\rm dof_{eff}}$ (based on eqn. 29 of \citep{2019PhRvD..99d3506R}). The asterisk sign indicates the parameter constraints which are sensitive to the choice of the priors. But as we can see from Figure. \ref{corner_sp}  $\log \rho_{\rm s}$ and $\log r_{\rm s}$ are degenerate among each other but do not correlate with the inferred value of the splashback radius. For $\log \rho_{\rm  o}$ see appendix \ref{sp_post}. }
    \label{modparam}
\end{table}

The location of the projected splashback radius $R_{\rm sp}$ was estimated using the minima of the logarithmic slope $d\log \xi_{\rm 2D}/ d\log R$ profile for the model predicted $\xi_{\rm 2D}$. The top right panel in Figure \ref{c-g-xcorr} shows the $d\log \xi_{\rm 2D}/ d\log R$ profile and the dark blue shaded regions represent the $68$ percent credible region around the median. In Table \ref{modparam} we present the median values for the model parameters and the location of the splashback radius along with errors computed from the 16-th and 84-th percentiles of the corresponding posterior distributions. The corresponding two-dimensional parameter posterior plots showing the correlation among different model parameters is presented in Appendix \ref{sp_post}. In our analysis, we obtain a $25$ percent constraint on the $R_{\rm sp}$ with a median at $1.19 \, h^{-1}{\rm  Mpc}$. The quantity $R_{\rm sp}$ corresponds to location of the steepest slope of the projected correlation function, and is expected to be at a location which is smaller than the location of the steepest slope in three dimensional distance from the cluster.
\begin{figure*}
    \centering
    \includegraphics[width=\textwidth]{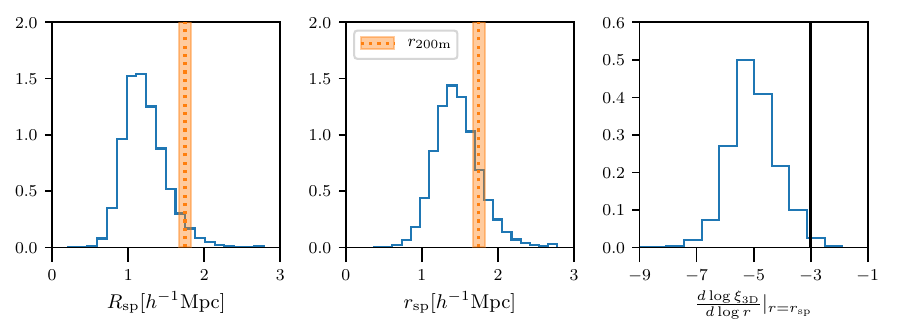}
    \caption{{\it Splashback radius posteriors:} The plot in the left and middle panels represents the distributions of the two-dimensional  $R_{\rm sp}$ and three-dimensional $r_{\rm sp}$ splashback radius values. The right panel shows the distribution of the logarithmic slope for the three-dimensional $\xi_{\rm 3D}(r)$ galaxy number density profile at $r_{\rm sp}$. The orange dotted line with a shaded region in each panel shows the weak lensing constraints on the spherical overdensity size $r_{\rm 200m}$. The solid black vertical line in the right panel denotes the expectation from the standard NFW profile.} 
    \label{xi-slpdist}
\end{figure*}

Therefore, we also infer the three-dimensional splashback radius $r_{\rm sp}$ for our cluster sample using our model. In Figure \ref{c-g-xcorr} the bottom left panel shows the three dimensional cross-correlation profile $\xi_{\rm 3D}(r)$ inferred from the model fits to the measurements of $\xi_{\rm 2D}(R)$, while the bottom right panel shows the corresponding logarithmic slope $d\log \xi_{\rm 3D}/ d\log r$. Here, the blue vertical dotted dash line with a light blue shaded region around it represents the location of splashback radius $r_{\rm sp }$. From our analysis we obtain $r_{\rm sp}=1.45 \pm 0.30 \,  h^{-1} {\rm Mpc}$. We note that the median values of $r_{\rm sp}$ and $R_{\rm sp}$ differ by roughly 20 percent as expected in cluster scale halos \citep[see e.g.,][]{2016ApJ...825...39M}.

Following \citet[][]{2017ApJ...841...18B}, in Figure \ref{xi-slpdist} we show the posterior distribution of the splashback radius in 2-d ($R_{\rm sp}$), and 3-d ($r_{\rm sp}$) along with the slope $ d\log \xi_{\rm 3D} / d\log r $ at the location of $r = r_{\rm sp}$. The value of the logarithmic slope we obtain $-5.05 \pm 0.88$ is steeper than $-3$ at more than $2\sigma$ (see Table~\ref{modparam}). The black solid vertical line denotes the logarithmic slope of $-3$ reached asymptotically on large scales for the NFW halos without the presence of the two-halo term. This shows that the location of the minimum corresponds to the splashback feature for our cluster sample associated with the steepest slope. We note that our slope value is consistent with those obtained in the X-ray study by \citet{2019MNRAS.485..408C} and \citet{2020PASJ...72...64M} study using optically selected clusters and also with values seen in the literature \citep{2016ApJ...825...39M, 2017ApJ...841...18B, 2017ApJ...836..231U, 2018ApJ...864...83C, 2019ApJ...874..184Z, 2019MNRAS.487.2900S, 2021ApJ...911..136B, 2021MNRAS.507.5758S}, especially given the large errors. These differences in the value of slope could possibly arise from the variations in the halo mass accretion rate in different samples, which correlates well with the sharpness in the splashback radius feature \citep{2014ApJ...789....1D} and more investigation will be done in future work.

We compare the values we infer for $r_{\rm sp}$ with the expectations from $\Lambda {\rm CDM}$ model. We estimate the approximate location for $r_{\rm sp}$ using the weak lensing calibrated halo mass estimate $M_{\rm 200m} = 10^{14.52} \, h^{-1} {\rm  M_\odot}$ at the median redshift of $z=0.46$ for our sample. We use the $M_{\rm 200m}$ and $z$ to computing the peak height $\nu_{\rm 200m} \equiv \delta_c/\sigma(M_{\rm 200m})/D(z)$. We then use the $\nu_{\rm 200m}$ as an input parameter in the fitting relations given by eqn. 7 of \citet[][]{2015ApJ...810...36M} and obtain  $r_{\rm sp} = 1.80 \pm 0.07\, h^{-1}{\rm  Mpc}$ as the  $\Lambda {\rm CDM}$ prediction. This estimate is shown by the solid black vertical line in Figure \ref{c-g-xcorr}, which is consistent with our results within $\sim 1.2 \sigma$. We further calculated the halo mass accretation rate $\Gamma =  5.86\pm 3.78$ for our sample using eqn 5 in \citet[][]{2015ApJ...810...36M}. The large error on the $\Gamma$ value is related to the broader constraints on the splashback feature $r_{\rm sp}$.

In the current analysis, we only work with a single z-band absolute magnitude cut $M_{z} -5\log h < -19.36$ which corresponds to an apparent magnitude limit of $m_{z} = 24.0$ at redshift $z = 0.75$. Note that the knee of the Schechter function is brighter in the z-band than in the i-band by about 0.3 magnitudes \citep{blanton_2001} . Our magnitude cut in the z-band is, therefore even fainter compared to the magnitude at the knee of the Schechter function than galaxies used in \citet[][]{2016ApJ...825...39M} who use an i-band magnitude cut of $M_{i} -5\log h < -19.43$. Such cuts are thus expected to minimize the effects of dynamical friction \citep[][]{2016JCAP...07..022A}. We have also checked the effects of dynamical friction using limits on flux which are two magnitudes brighter limits similar to tests commonly carried out in the literature \citep[for, e.g.][]{2016ApJ...825...39M, 2020PASJ...72...64M}. We found no significant change in the splashback feature, given the large errors resulting from the low signal-to-noise cross-correlation measurements. We intend to pursue such studies in more detail in future work.

\begin{figure}
    \centering
    \includegraphics[width=\columnwidth]{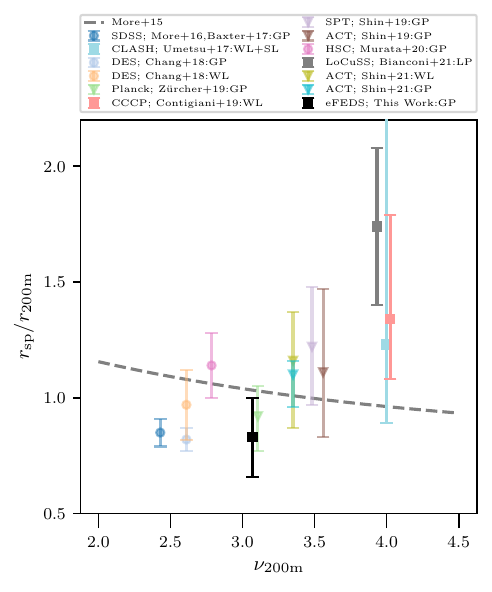}
    \caption{{\it Comparison with other studies:} In the above figure, we compare our results with the existing works in the literature. The x-axis represents the peak height $\nu_{\rm 200m}$ and the y-axis represents the ratio between splashback radius $r_{\rm sp}$ and the spherical overdensity $r_{\rm 200m}$ radius estimates. The round data points show the values computed using the optically selected clusters, the triangular points are for the analysis using SZ selected clusters, and the squared points are for the X-ray selected clusters. In the legends, we label each datapoint using their survey name, reference and the analysis method used - the weak lensing (WL), strong lensing (SL), galaxy number density (GP) or stacked Luminosity profile (LP). The dashed grey line shows the expectation from the $\Lambda$CDM cosmology as a function of $\nu_{\rm 200m}$ computed using the fitting function given by \citet{2015ApJ...810...36M}. The black datapoint with errors represents our results from this study using the eFEDS X-ray clusters and shows marginal consistency ($1.2\sigma$) with the $\Lambda$CDM predictions.     }
    \label{cmp_res}
\end{figure}

Finally, in Figure \ref{cmp_res} we compare our results with those in the literature in the $\nu_{\rm 200m}-r_{\rm sp}/r_{\rm 200m}$ plane. Given that this relation is fairly independent of redshift \citep{2015ApJ...810...36M}, this allows for a uniform comparison of results in the literature. The grey dashed line represents the expectations from $\Lambda{\rm CDM}$ using the parameterization given by eqn. 7 in \citet[][]{2015ApJ...810...36M} and calibrated from numerical simulations. Round data points are used to present the constraints from optically selected clusters, triangular data points for those obtained using clusters identified using the SZ effect, and square data points for those obtained using X-ray selected clusters. In the legend, we have also labelled the name of the data used along with the method used in the corresponding investigation - galaxy number density profile (GP), weak lensing (WL) and stacked luminosity profile (LP). The black square point shows our constraint based on the galaxy number density profile around eFEDS X-ray clusters from this work, which yields $r_{\rm sp}/ r_{\rm 200m} \approx 0.83 \pm 0.17$. This results is a 20 percent constraint and yields results consistent $\approx 1.2\sigma$ with the $\Lambda {\rm CDM}$ prediction. It represents an improvement over the weak-and-strong lensing based study on CLASH X-ray cluster sample by \citet{2017ApJ...836..231U} which provided a lower limit for $r_{\rm sp}/ r_{\rm 200m} > 0.89$. Similarly, our analysis also puts nearly 25 percent tighter bounds than the one obtained with another weak lensing result using the CCCP X-ray clusters \citep{2019MNRAS.485..408C}.

We find that our constraints are similar to those obtained by galaxy number density studies using SZ clusters by \citet[][]{2019ApJ...874..184Z, 2019MNRAS.487.2900S}, the weak lensing-based analysis in \citet[][]{2021MNRAS.507.5758S} using ACT SZ clusters and the stacked luminosity profile results in LoCuSS X-ray clusters \citep[][]{2021ApJ...911..136B}. The precision with which we determine the splashback radius is worse when compared to galaxy density profile studies using SZ selected clusters \citep[][]{2021MNRAS.507.5758S} or optically selected clusters \citep[][]{2016ApJ...825...39M, 2017ApJ...841...18B, 2018ApJ...864...83C, 2020PASJ...72...64M}, albeit the latter are likely affected by systematics related to projection effects. This is expected to get better as our sample size gets larger.

\section{Conclusions}
\label{conclusion}
The splashback radius, which denotes the physical boundary of dark matter halos, depends upon the current accretion mass rate. Various studies in the literature have attempted to constrain the location of this radius in observations. In this work, we studied the splashback radius around 109 eROSITA eFEDS X-ray selected clusters by cross-correlating them with the HSC S19A optical photometric galaxies. Our use of X-ray selected clusters avoids systematics related to projection effects which have affected the measurement of the splashback radius from optically selected clusters. We select X-ray clusters having luminosities above a threshold $L_X > 10^{43.5} {\rm erg \,s^{-1} }h^{-2}$ within the redshift $z<0.75$, which provides us a nearly volume limited sample. Finally, we compared our inferred value of the splashback radius with the standard spherical overdensity radius $r_{\rm 200m}$ calibrated using weak gravitational lensing. 

We briefly summarize our major findings below :
\begin{itemize}
    \item We measure the stacked weak lensing signal around our cluster sample using the background galaxies from HSC S16A shape catalogue data and obtain measurements with a signal-to-noise of 17.93. We then use a simple NFW profile to model the stacked signal and infer a  halo mass $\log [M_{\rm 200m}/h^{-1}{\rm M_\odot}] = 14.52 \pm 0.06 $ which corresponds to an spherical overdensity size of $r_{\rm 200m} = 1.75 \pm 0.08\,h^{-1}{\rm  Mpc}$.

    \item We measure the projected cross-correlation with a signal-to-noise of 17.43 for our X-ray galaxy clusters using the optical galaxies from HSC S19A data having z-band absolute magnitude cut $M_{z} -5\log h < -19.36$. We model these measurements by projecting the three-dimensional functional form given by \citet{2014ApJ...789....1D} to infer the location of the steepest slope and associate it with the splashback radius.
   
    \item We find the value for the 3D steepest slope to be $-5^{+0.88}_{-0.73}$, more than $2\sigma$ away from the asymptotic $-3$ value in the case of the standard NFW  profile which provides evidence for the presence of a splashback feature. 
  
    \item Our analysis corresponds to $\sim 25$ percent constraint on the projected $R_{\rm sp} = 1.19^{+0.30}_{-0.22}\, h^{-1}{\rm  Mpc}$ and $\sim 20$ percent constraint on the three dimensional $r_{\rm sp} = 1.45^{+0.30}_{-0.26} \, h^{-1}{\rm  Mpc}$ splashback radius. These values are in the range as expected commonly for the massive halos in numerical simulations \citep{2014ApJ...789....1D, 2015ApJ...810...36M}.
    
    \item Our constraints on $r_{\rm sp }$ are broadly comparable to spherical overdensity estimates $r_{\rm 200m} =1.75\pm 0.08 \, h^{-1}{\rm  Mpc}$ and are marginally consistent ($\approx 1.2 \sigma$) with the expectation from numerical simulation for the weak lensing calibrated halo mass at the median redshift $z=0.46$ for our cluster sample.  The results from our analysis significantly improve the errors on the splashback radius measurement based on X-ray selected galaxy clusters.
    
    \item We infer a halo mass accretion rate $\Gamma =5.86 \pm 3.78$ with large errors, where the error is dominated by the error on the inferred splashback radius $r_{\rm sp}$. 
    
\end{itemize}
In our analysis, we use X-ray centres of the galaxy clusters for the weak lensing and cross-correlation signal measurements. We have checked the dependence of our results on the choice of the centre by using the optically confirmed galaxy as the centre and find that our results are robust to this choice. Given the differences between X-ray and optical centres - $92.6^{+44.3}_{-35.1} {\rm kpc}$ \citep[][]{2022arXiv221210107_Seppi}, we have also explore effects of miscentering on our conclusions. We have reanalyzed both the weak lensing and cross-correlation measurements after removing signals in radial bins below $200\, h^{-1}{\rm kpc}$ and found no significant change in our results. We have also checked the effect of the inclusion of $k$ and $n_{H}$ corrections on the clusters for our measurements by using X-ray cluster data from the catalogue given by \citet[][]{2022A&A...661A...2L}, which changes our sample by about 10 percent. We re-run the projected galaxy number density and weak lensing analysis for the changed sample. We found a value of $r_{\rm sp}/r_{\rm 200m} \approx 0.73\pm 0.24$, which differs by about 12 per cent and is within the 25 per cent error of our fiducial measurement. Thus our results are fairly robust to changes in our sample selection.

We also want to point out that in the present analysis, we only use one z-band absolute limit on the S19A optical galaxies for cross-correlation measurements. We choose the faintest possible limit such that we can get less noisy measurements on large scales. In future studies with a larger cluster sample, we would also like to test for possible dynamical friction effects.

Our results provide meaningful constraints on the splashback radius of galaxy clusters, although with large statistical errors. However, in the near future, we expect better constraints by using all-sky cluster catalogues from the eRASS \citep[][]{2012arXiv1209.3114M}. The statistical precision of these measurements will allow comparisons with expectations from the hydrodynamic simulations like IllustrisTNG \citep[][]{2017MNRAS.465.3291W,2018MNRAS.473.4077P} and thorough investigations of any differences between the observations and theoretical predictions. It will also allow us to explore the dependence of the splashback feature on X-ray cluster observables such as dynamical state, X-ray morphology and hydrostatic halo masses.

\section*{Acknowledgements}
We are grateful to  Navin Chaurasiya, Amit Kumar, Moun Meenakshi, Preetish K. Mishra, Ayan Mitra, Masahiro Takada and Keiichi Umetsu for discussions and their insightful comments on the earlier version of the manuscript. DR thank the University Grants Commission (UGC) of India, for providing financial support as a senior research fellow. We acknowledge the use of  the high performance computing facility - Pegasus at IUCAA. This work was supported in part by the World Premier International Research Center Initiative (WPI Initiative), MEXT, Japan, and JSPS KAKENHI Grant Nos. 20H01932 and 21H05456. E.B. acknowledges financial support from the European Research Council (ERC) Consolidator Grant under the European Union’s Horizon 2020 research and innovation programme (grant agreement CoG DarkQuest No 101002585).

The Hyper Suprime-Cam (HSC) collaboration includes the astronomical communities of Japan and Taiwan, and Princeton University. The HSC instrumentation and software were developed by
the National Astronomical Observatory of Japan (NAOJ), the Kavli
Institute for the Physics and Mathematics of the Universe (Kavli
IPMU), the University of Tokyo, the High Energy Accelerator Research Organization (KEK), the Academia Sinica Institute for Astronomy and Astrophysics in Taiwan (ASIAA), and Princeton University. Funding was contributed by the FIRST program from the
Japanese Cabinet Office, the Ministry of Education, Culture, Sports,
Science and Technology (MEXT), the Japan Society for the Promotion of Science (JSPS), Japan Science and Technology Agency
(JST), the Toray Science Foundation, NAOJ, Kavli IPMU, KEK,
ASIAA, and Princeton University.

This paper is based in part on data collected at the Subaru
Telescope and retrieved from the HSC data archive system, which is
operated by Subaru Telescope and Astronomy Data Center (ADC)
at NAOJ. Data analysis was in part carried out with the cooperation
of Center for Computational Astrophysics (CfCA), NAOJ.

This work is based on data from eROSITA, the soft X-ray instrument aboard SRG, a joint Russian-German science mission supported by the Russian Space Agency (Roskosmos), in the interests of the Russian Academy of Sciences represented by its Space Research Institute (IKI), and the Deutsches Zentrum f\"{u}r Luft- und Raumfahrt (DLR). The SRG spacecraft was built by Lavochkin Association (NPOL) and its subcontractors, and is operated by NPOL with support from the Max Planck Institute for Extraterrestrial Physics (MPE). The development and construction of the eROSITA X-ray instrument was led by MPE, with contributions from the Dr. Karl Remeis Observatory Bamberg \& ECAP (FAU Erlangen-Nuernberg), the University of Hamburg Observatory, the Leibniz Institute for Astrophysics Potsdam (AIP), and the Institute for Astronomy and Astrophysics of the University of T\"{u}bingen, with the support of DLR and the Max Planck Society. The Argelander Institute for Astronomy of the University of Bonn and the Ludwig Maximilians Universit\"{a}t Munich also participated in the science preparation for eROSITA. 


\section*{Data Availability}
This work uses publically available catalogue data accessible through survey websites. The eROSITA eFEDS X-ray cluster catalogue is available at \href{https://erosita.mpe.mpg.de/edr/index.php}{https://erosita.mpe.mpg.de/edr/index.php} and Subaru HSC dataset can be found at \href{https://hsc-release.mtk.nao.ac.jp/doc/}{https://hsc-release.mtk.nao.ac.jp/doc/}. The measurements for cross-correlation and weak lensing analysis are available on \href{https://github.com/divyarana-cosmo/rsp\_efeds\_hsc\_s19a}{https://github.com/divyarana-cosmo/rsp\_efeds\_hsc\_s19a}.




\bibliographystyle{mnras}
\bibliography{ms} 



\appendix

\section{Boost Parameters}
\label{apx_boost}
\begin{figure}
    \centering
    \includegraphics[width=\columnwidth]{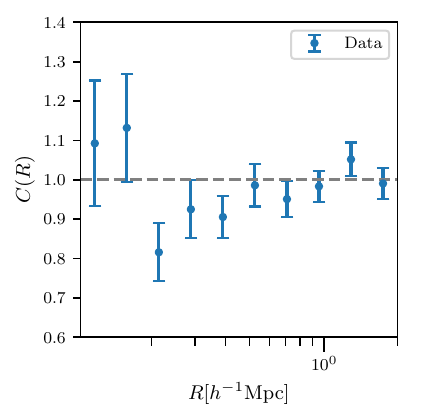}
    \caption{{\it Boost parameters:} The blue data points denotes the boost parameter $C(R)$ for our weak lensing measurement $\Delta\Sigma$ and are estimated using eqn. \ref{boost_eqn.} with y errors computed from the scatter in random realizations. The plot shows consistency of boost parameter with grey dashed unity line.}
    \label{boost}
\end{figure}

As discussed in Section \ref{sec-wl}, we estimate the boost parameters using eqn. \ref{boost_eqn.} to quantify the dilution in the weak lensing signal due to the systematics in the photometric redshift $p(z)$ distribution in the source galaxies. In Figure \ref{boost}, the blue data points represent the boost parameters from our weak lensing signal with $1\sigma$ error bars from random realizations. We found that they are mostly consistent with the unity line apart from the third datapoint, which differs by more than $2\sigma$. We test the impact of the third data point on our inferred halo mass. We reran our signal fitting after removing the third data point and found no significant change in the inferred value of the halo mass.

\section{Covariances}
\label{meascov}
\begin{figure}
    \centering
    \includegraphics[width=\columnwidth]{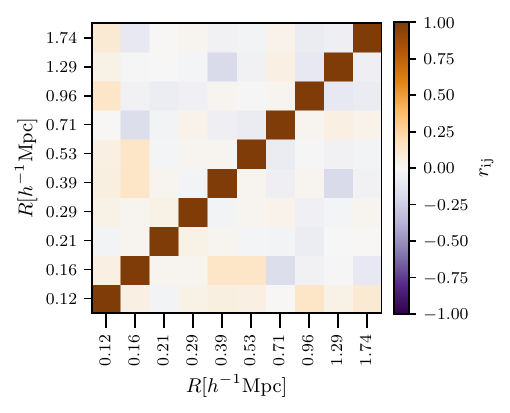}
    \caption{{\it Weak lensing profile covariance:} The above plot shows the correlation coefficient $r_{\rm ij}$ matrix for the $i^{\rm th}$ and $j^{\rm th}$ radial bin of the weak lensing signal measurements $\Delta \Sigma$ given in Sec \ref{sec-wl} and estimated by shape noise using 200 different random rotations of background source galaxies with x-y axis showing our radial binning for signal measurements.}
    \label{wl_cov}
\end{figure}

\begin{figure}
    \centering
    \includegraphics[width=\columnwidth]{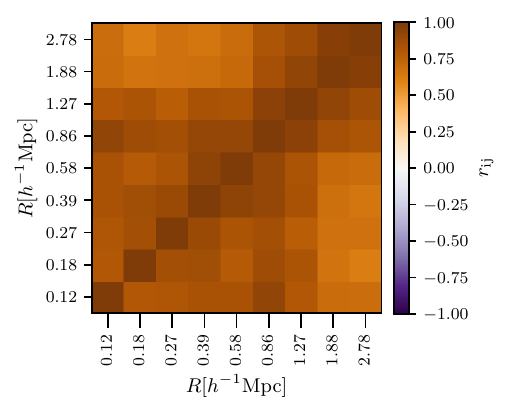}
    \caption{{\it Galaxy number density covariance:} The above plot present the correlation coefficient $r_{\rm ij}$ matrix for the $i^{\rm th}$ and $j^{\rm th}$ radial bin of our galaxy number density profile measurements $\xi_{\rm 2D}$ given in Sec \ref{sec-c-g-xcorr} and estimated using 20 jackknife region with each having roughly $5\, {\rm deg^2}$ area.The x and y axis denotes our radial binning for signal measurements.}
    \label{sp_cov}
\end{figure}

Figure \ref{wl_cov} and Figure \ref{sp_cov} show the correlation coefficient $r_{\rm ij}$ for the corresponding covariances of the weak lensing and galaxy density profile measurements. We estimate $r_{\rm ij}$ from the covariance $C$ and given by

\begin{equation}
    r_{\rm ij} = \frac{C_{\rm ij}}{\sqrt{C_{\rm ii} C_{\rm jj}}}
\end{equation}

where subscript ${\rm ij}$ represents the ${\rm i}^{\rm th}$ and ${\rm j}^{\rm th}$ radial bins. We use the shape noise covariance for the weak lensing profile and area jackknifes to compute covariance for the galaxy number density profile. We describe the details in Sec \ref{measurements}.

Figure \ref{wl_cov} shows mostly no correlation between the weak lensing signal among different radial bins, as it is dominated by the shape noise in the scales of our interest. The source galaxies that contribute to the signal in each radial bin are different and thus independent. At the same time, we found a mild positive correlation for the galaxy number density measurement shown by the off-diagonal elements in Figure \ref{sp_cov}. This is consistent with correlations in the small scale clustering of satellite galaxies around clusters. These correlations arise as an increase in the satellite numbers in smaller radial bins typically also indicate a similar variation at larger radii in a cluster.

\section{Degeneracies in the model parameters}
\begin{figure*}
    \centering
    \includegraphics[width=\textwidth]{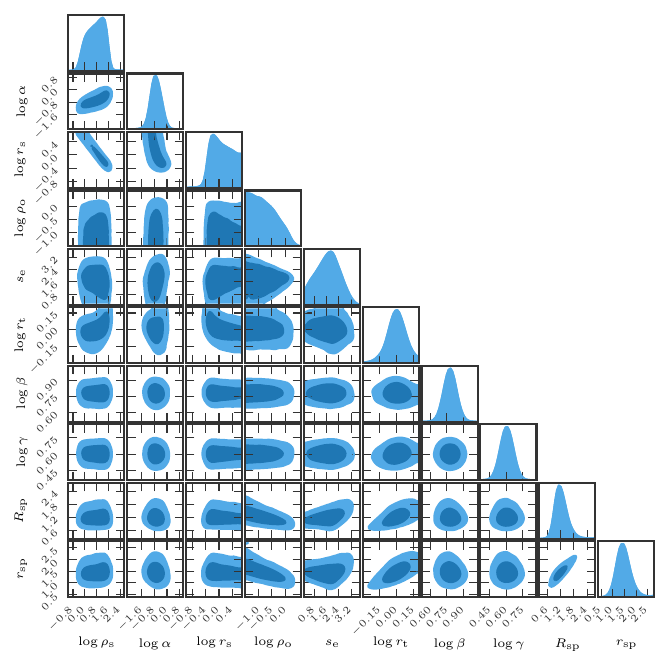}
    \caption{{\it Splashback parameter posteriors:} The above corner plot shows the degeneracies among the model parameters described in Sec \ref{sec-c-g-xcorr} for the projected cluster-galaxy cross-correlation measurements. }
    \label{corner_sp}
\end{figure*}

Figure \ref{corner_sp} shows the correlations between the parameters used for modelling the cluster-galaxy cross-correlation measurements. The parametric dependence is given by the function form in eqn. \ref{dk14_in}, \ref{dk14_out} and \ref{dk14_trans}. We note that the posterior sample for $\log {\rho_{\rm o}}$ pile up at the lower edge of the prior range. We tested our results with broader priors and found some values for $R_{\rm sp}$ to be at the largest radial bins. These values arise from the poor signal measurements at large scales to get a well-constrained outer profile. In such cases, the model tries to fit the whole measurement without the outer profile by using a smaller amplitude $\log {\rho_{ \rm o}}$ giving rise to a spurious secondary peak in the $R_{\rm sp}$ posterior. 

\label{sp_post}


\bsp	
\label{lastpage}
\end{document}